\documentstyle[12pt,epsfig,pstricks]{article}
\catcode`\@=11
\long\def\@makefntext#1{
\protect\noindent \hbox to 3.2pt {\hskip-.9pt  
$^{{\ninerm\@thefnmark}}$\hfil}#1\hfill}                %CAN BE USED 

\def\@makefnmark{\hbox to 0pt{$^{\@thefnmark}$\hss}}  %ORIGINAL 
        
\def\ps@myheadings{\let\@mkboth\@gobbletwo
\def\@oddhead{\hbox{}
\rightmark\hfil\ninerm\thepage}   
\def\@oddfoot{}\def\@evenhead{\ninerm\thepage\hfil
\leftmark\hbox{}}\def\@evenfoot{}
\def\sectionmark##1{}\def\subsectionmark##1{}}

\renewcommand{\thefootnote}{\fnsymbol{footnote}}

\newcounter{sectionc}\newcounter{subsectionc}\newcounter{subsubsectionc}
\renewcommand{\section}[1] {\vspace*{0.6cm}\addtocounter{sectionc}{1} 
\setcounter{subsectionc}{0}\setcounter{subsubsectionc}{0}\noindent 
        {\normalsize\bf\thesectionc. #1}\par\vspace*{0.4cm}}
\renewcommand{\subsection}[1] {\vspace*{0.6cm}\addtocounter{subsectionc}{1} 
        \setcounter{subsubsectionc}{0}\noindent 
        {\normalsize\it\thesectionc.\thesubsectionc. #1}\par\vspace*{0.4cm}}
\renewcommand{\subsubsection}[1]
{\vspace*{0.6cm}\addtocounter{subsubsectionc}{1}
        \noindent {\normalsize\rm\thesectionc.\thesubsectionc.\thesubsubsectionc. 
        #1}\par\vspace*{0.4cm}}

\newcounter{appendixc}
\newcounter{subappendixc}[appendixc]
\newcounter{subsubappendixc}[subappendixc]

\renewcommand{\appendix}[1] {\vspace*{0.6cm}
        \refstepcounter{appendixc}
        \setcounter{figure}{0}
        \setcounter{table}{0}
        \setcounter{equation}{0}
        \renewcommand{\thefigure}{\Alph{appendixc}.\arabic{figure}}
        \renewcommand{\thetable}{\Alph{appendixc}.\arabic{table}}
        \renewcommand{\theappendixc}{\Alph{appendixc}}
        \renewcommand{\theequation}{\Alph{appendixc}.\arabic{equation}}
        \noindent{\bf Appendix \theappendixc #1}\par\vspace*{0.4cm}}

\def\abstracts#1{{
        \centering{\begin{minipage}{12.2truecm}\footnotesize\baselineskip=12pt\noindent
        \centerline{\footnotesize ABSTRACT}\vspace*{0.3cm}
        \parindent=0pt #1
        \end{minipage}}\par}} 

\renewenvironment{thebibliography}[1]
        {\begin{list}{\arabic{enumi}.}
        {\usecounter{enumi}\setlength{\parsep}{0pt}
\setlength{\leftmargin 1.25cm}{\rightmargin 0pt}
         \setlength{\itemsep}{0pt} \settowidth
        {\labelwidth}{#1.}\sloppy}}{\end{list}}
\def\Journal#1#2#3#4{{#1} {\bf #2}, #3 (#4)}
 
\def\NIMA{{\em Nucl. Instrum. Methods} A}
\def\NPB{{\em Nucl. Phys.} B}
\def\PLB{{\em Phys. Lett.}  B}

\def\PRD{{\em Phys. Rev.} D}
\def\ZPC{{\em Z. Phys.} C}
\def\EPJ{{\em Eur. Phys. J.} C}

\newcounter{itemlistc}
\newcounter{romanlistc}
\newcounter{alphlistc}
\newcounter{arabiclistc}

\newcommand{\fcaption}[1]{
        \refstepcounter{figure}
        \setbox\@tempboxa = \hbox{\footnotesize Fig.~\thefigure. #1}
        \ifdim \wd\@tempboxa > 6in
           {\begin{center}
        \parbox{6in}{\footnotesize\baselineskip=12pt Fig.~\thefigure. #1}
            \end{center}}
        \else
             {\begin{center}
             {\footnotesize Fig.~\thefigure. #1}
              \end{center}}
        \fi}

\newcommand{\tcaption}[1]{
        \refstepcounter{table}
        \setbox\@tempboxa = \hbox{\footnotesize Table~\thetable. #1}
        \ifdim \wd\@tempboxa > 6in
           {\begin{center}
        \parbox{6in}{\footnotesize\baselineskip=12pt Table~\thetable. #1}
            \end{center}}
        \else
             {\begin{center}
             {\footnotesize Table~\thetable. #1}
              \end{center}}
        \fi}

\def\@citex[#1]#2{\if@filesw\immediate\write\@auxout
        {\string\citation{#2}}\fi
\def\@citea{}\@cite{\@for\@citeb:=#2\do
        {\@citea\def\@citea{,}\@ifundefined
        {b@\@citeb}{{\bf ?}\@warning
        {Citation `\@citeb' on page \thepage \space undefined}}
        {\csname b@\@citeb\endcsname}}}{#1}}

\newif\if@cghi
\def\cite{\@cghitrue\@ifnextchar [{\@tempswatrue
        \@citex}{\@tempswafalse\@citex[]}}
\def\citelow{\@cghifalse\@ifnextchar [{\@tempswatrue
        \@citex}{\@tempswafalse\@citex[]}}
\def\@cite#1#2{{$\null^{#1}$\if@tempswa\typeout
        {IJCGA warning: optional citation argument 
        ignored: `#2'} \fi}}

 1
 1
 1

\font\ninerm=cmr9

\begin{document}
\begin{flushright}
{\bf PRA-HEP/98-05} \\
July 1998
\end{flushright}
\vspace*{1.5cm}
%\centerline{\normalsize\bf \Large Jets as a Source of Information About Photon 
%Structure \footnote{To appear in the {\it Proceedings of the DIS98
%Workshop}, held in Brussels, Belgium, April 4 - 8, 1998.}}
\centerline{\normalsize\bf JETS AS A SOURCE OF INFORMATION}
\centerline{\normalsize\bf ABOUT PHOTON STRUCTURE \footnote{To appear in the 
{\it Proceedings of the DIS98 Workshop}, held in Brussels, Belgium, 
April 4 - 8, 1998.}}
\vspace*{0.6cm}
\centerline{\footnotesize MAREK TA\v{S}EVSK\'{Y}}
\baselineskip=13pt
\centerline{\footnotesize\it Institute of Physics, Academy of Sciences
of the Czech Republic, Na Slovance 2}
\baselineskip=12pt
\centerline{\footnotesize\it Praha, CZ - 182 21, Czech Republic}
\centerline{\footnotesize E-mail: tasevsky@fzu.cz}
\vspace*{0.3cm}
\centerline{\footnotesize On behalf of the H1 Collaboration.}

\vspace*{0.9cm}
\abstracts{A review of recent jet measurements of the photon 
structure from the H1 experiment at HERA is presented. A manifestation of the
photon structure in single-jet and di-jet cross sections is shown. From di-jet
cross sections the effective parton distribution function is extracted and its
dependence on the scale as well as the photon virtuality is discussed.}
%The scale and the virtuality dependence of the effective parton distribution
%function of the photon is discussed.}
 
%\vspace*{0.6cm}
\normalsize\baselineskip=15pt
\setcounter{footnote}{0}
\renewcommand{\thefootnote}{\alph{footnote}}
\section{Introduction}
The photon, as an elementary particle, is considered to be structureless. 
And yet, from numerous experimental data it follows that in high energy 
$\gamma \gamma$ or $\gamma p$ interactions the photon behaves like a hadron, 
an object with a structure. Quantum mechanically, the photon can fluctuate 
into a fermion-antifermion pair which can further evolve in the framework 
of QED and QCD. This is called photon structure and it can be investigated 
by means of the jets. Like those of the proton, parton distribution functions 
(PDF) of the photon depend on the probing scale\cite{rick}, but additionally 
also on its virtuality $Q^2$. This latter dependence was measured in the 
single-jet study\cite{tania} and more recently in an H1 di-jet analysis 
where the effective PDF of the virtual photon was extracted for the first time.

\section{Single-Jet and Di-Jet Cross Sections}
For appropriately normalised matrix elements $M_{ij}^{kl}$ and 
neglecting the contribution of the longitudinally polarised photons the
expression for the resolved photon contribution to the di-jet cross section
in $ep$ collisions reads
\begin{equation}
\frac{\mathrm{d}\sigma^{ep}}{{\mathrm{d}y} {\mathrm{d}}x_\gamma 
{\mathrm{d}}x_p 
{\mathrm{d}}\!\cos\!\theta {\mathrm{d}}Q^{2}}=
%\frac{1}{32 \pi s_{ep}}
\frac{f_{\gamma/e}(y,Q^2)}{y}
\sum\limits_{i,j,k,l} \frac{f_{i/\gamma}(x_\gamma,\mu_F^2,Q^2)}{x_\gamma}
\frac{f_{j/p}(x_p,\mu_F^2)}{x_p}
|M_{ij}^{kl}|^2
\label{eq:gener}
\end{equation}
where $f_{\gamma/e}$ denotes the flux of transversely polarised virtual
photons inside electron, $f_{i/\gamma},f_{j/p}$ are PDF of the photon and 
proton respectively taken at $x_\gamma$($x_p$), the fraction of the 
photon(proton) momentum carried by the parton entering the hard subprocess,
and at the factorization scale $\mu_F$.
The $M_{ij}^{kl}$ are the leading order (LO) matrix elements of the binary 
parton level processes $ij\rightarrow kl$, proportional to $\alpha_s(\mu_R^2)$.
We assumed $\mu_F=\mu_R=E_T$ (the transverse energy of the jet).
 
To correct for detector, fragmentation and higher order effects, PYTHIA and 
HERWIG MC event generators were used for photoproduction 
and HERWIG, RAPGAP, ARIADNE and LEPTO for low $Q^2$ events.
As input PDFs of the virtual photon, either the simple Drees-Godbole\cite{DG} 
(DG) virtuality suppression factor 
\begin{equation}
L(E_T^2,Q^2,\omega) = 
\frac{\ln((E_T^2+\omega^2)/(Q^2+\omega^2))}{\ln((E_T^2+\omega^2)/\omega^2))},
\;\;{\rm with\;free\;parameter}\; \omega 
\end{equation}
multiplying the PDFs of the real photon (as $L$ for quarks and $L^2$ for 
gluons) or set D of the SaS\cite{SaS} parameterisations was taken.  
 
In Fig.1a the photoproduction di-jet cross sections are compared with
next-to-leading order (NLO) calculations\cite{KK} in which the photon PDFs 
were taken from $F_2^\gamma$ measurements in $\gamma \gamma$ interactions 
at LEP (and also from PEP, PETRA and TRISTAN data).
The data were corrected for detector effects by an unfolding 
procedure\cite{unfold}.
The satisfactory overall description of the di-jet $ep$ data supports the 
universality of the photon parton distributions. However, the data indicate
that adjustments of the PDF at high $x_\gamma$ are required where no 
$F_2^{\gamma}$ measurements have been made so far.
 
In Fig.1b the inclusive single-jet cross sections as a function  
of $Q^2$ are compared to predictions of LEPTO which contains only
the direct coupling of the photon to quarks. The prediction agrees with the 
data well in the DIS region where, typically, $E_T^2 \ll Q^2$. However, the
model cannot describe the data in the region $E_T^2 > Q^2$ where the 
contribution of the virtual resolved photon is expected to be important. 
If the same data are compared to HERWIG and RAPGAP, generators 
containing resolved processes, a good agreement is achieved only if the photon 
PDF is suppressed with increasing $Q^2$ \cite{tania}.
%The generator ARIADNE (Fig.2) effectively has a resolved photon component due
%to its colour dipole mechanism which describes however the jet cross sections
%only at large $Q^2$. 
The di-jet data, on the other hand, enable to construct the distribution
of $x_\gamma$ --- a quantity closely related to the partonic content of the
photon --- or more, to extract the effective parton distribution function of 
the photon (see section 3). In the more recent H1 di-jet analysis from which
preliminary plots 2, 3a, 4 are shown, the most important cuts imposed on the
data were: $E_T > 5$ GeV and $-3 < \eta < 0\;$ for two highest $E_T$ jets 
constructed on clusters and tracks in $\gamma^* p$ cms using CDF 
cone algorithm\cite{CDF}. The results were integrated over the region of inelasticity,
$0.25 < y <0.7$, and studied in the region of the photon virtuality, 
$1.4 < Q^2 < 25$ GeV$^2$. The limits on $E_T$ and $Q^2$ were driven by the
condition $E_T^2 \gg Q^2$ --- a natural requirement when investigating
the virtual photon structure (see the mean values of $E_T^2$ and $Q^2$ in 
Fig.4). The data were corrected using the mentioned unfolding 
procedure\cite{unfold}.
The virtuality dependence of the di-jet cross sections, 
$\mathrm{d}\sigma/\mathrm
{d}x_\gamma$, is shown in Fig.2. Again, the HERWIG and RAPGAP 
models incorporating a resolved photon contribution with a virtuality-dependent
suppression, describe the data satisfactorily. In Fig.3a, the ratio of 
resolved to direct contributions, defined as 
$\sigma(x_\gamma \le 0.75)$/$\sigma(x_\gamma>0.75)$ is plotted as a function 
of $Q^2$. The data in Figs.2 and 3a are compatible with DG virtuality 
suppression for $0.05\le \omega\le 0.1$ GeV.

\section{Effective Parton Distribution of the Photon}
The concept of effective PDF was developed for jet analyses at CERN 
SPS\cite{ses}
where it was not possible to distinguish the contributions of individual
subprocesses $ij\rightarrow kl$ to the sum in Eq.~(\ref{eq:gener}). 
Applied to $ep$ collisions at HERA, this approach allows
us to approximate the formula (\ref{eq:gener}) by a simple product 
\begin{equation}
\frac{\mathrm{d}\sigma^{ep}}{{\mathrm{d}y} {\mathrm{d}}x_\gamma 
{\mathrm{d}}x_p 
{\mathrm{d}}\!\cos\!\theta {\mathrm{d}}Q^{2}} \approx
\frac{f_{\gamma/e}(y,Q^2)}{y}
\frac{f^{\gamma}_{\mathrm{eff}}(x_\gamma,E_T^2,Q^2)}{x_\gamma}
\frac{f^{p}_{\mathrm{eff}}(x_p,E_T^2)}{x_p}
|M_{SES}|^2
\end{equation}
of a {\em single effective subprocess} matrix element $M_{SES}$ and the 
{\em effective PDF} of the colliding photon and proton 
\begin{equation}
f_{\mathrm{eff}}^{\gamma/p}(x,E_T^2,Q^2) =
\sum\limits_i^{n_f} [q_i(x,E_T^2,Q^2) + \overline{q}_i(x,E_T^2,Q^2)] +
\frac{9}{4} g(x,E_T^2,Q^2).
\label{eq:effpdf}
\end{equation}

The measured di-jet cross sections were corrected to the LO di-parton
cross section by the unfolding procedure\cite{unfold}. The effective PDF of 
the photon $f_{\mathrm{eff}}^{\gamma}$ was then determined by comparing the 
measured di-parton cross section
with that calculated by MC model using the GRV-LO photon PDF and incorporating,
for the virtual photon, the DG suppression factor: 
\begin{equation}
f_{{\mathrm eff}}^{\gamma, \mathrm DATA}(x_\gamma,E_T^2,Q^2) = f_{{\mathrm
eff}}^{\gamma,\mathrm GRV-LO} (x_\gamma,E_T^2,Q^2)
L(E_T^2,Q^2,\omega^2)
\frac{{\mathrm d}\sigma^{\mathrm DATA}/{\mathrm d}x_{\gamma}}
{{\mathrm d}\sigma^{\mathrm MC,GRV-LO}/{\mathrm d}x_{\gamma}}
\label{eq:effdat}
\end{equation}
In Fig.3b the effective PDF
of the (almost) real photon integrated over the range $0.4 < x_\gamma < 0.7$
is shown as a function of the scale $E_T^2$. In this $x_{\gamma}$
region $f_{\mathrm eff}^{\gamma}$ is dominated by the quarks. 
Also shown are the PDFs from
 GRV-LO parameterisations of the photon\cite{grv-photon}, which is in good 
agreement with the data, and of the pion\cite{grv-pion}, which differs from 
measurement both in shape and absolute rate.
The increase of $f_{\mathrm eff}^{\gamma}$ with $E_T^2$ seen in the data
is compatible with a logarithmic dependence of $q^{\gamma}$ on $\mu_F$, which 
is expected from the inhomogeneous term in the DGLAP evolution equation.   
In Fig.4 the effective PDF of the
virtual photon is presented in four $Q^2$ intervals and compared 
with the predictions of GRV--LO 
(for $\omega = 0.05$ and 0.1 GeV)
and with SaS1D and SaS2D parameterisations. The data are in good agreement with
the GRV for both values of $\omega$, the SaS 
parameterisations tend to fall more rapidly in the low $x_\gamma$ region than 
the data.
 
\section{Conclusion}
New kinematical regions were explored by the H1 experiment in single-jet and 
di-jet analyses. From photoproduction data the scale dependence of the photon
effective PDF was determined. The observed logarithmic increase agrees 
with the prediction of perturbative QCD.  The low $Q^2$ data indicate
the importance of the resolved photon contribution to the 
total jet cross section and its suppression with increasing virtuality.   
The $Q^2$ dependence of the di-jet cross section and of the ratio
$\sigma^{res}/\sigma^{dir}$ was shown.
The first measurements of the LO effective PDF of the photon in the region 
 $1.4 < Q^2 <25$ GeV$^2$ was presented. 

\section{Acknowledgements}
I thank J. Ch\'{y}la and S. Maxfield for the reading of the text
and for comments.
This work is supported by the Grant Agency of Academy of 
Sciences of the Czech Republic under the grant no. A1010821,
by the Grant Agency of the Czech Republic under the 
grant no. 202/96/0214 and by the Grant Agency of Charles University under 
the grant number 177.

\newpage
\begin{minipage}{72mm}
\hspace*{-0.2cm}
\epsfig{file=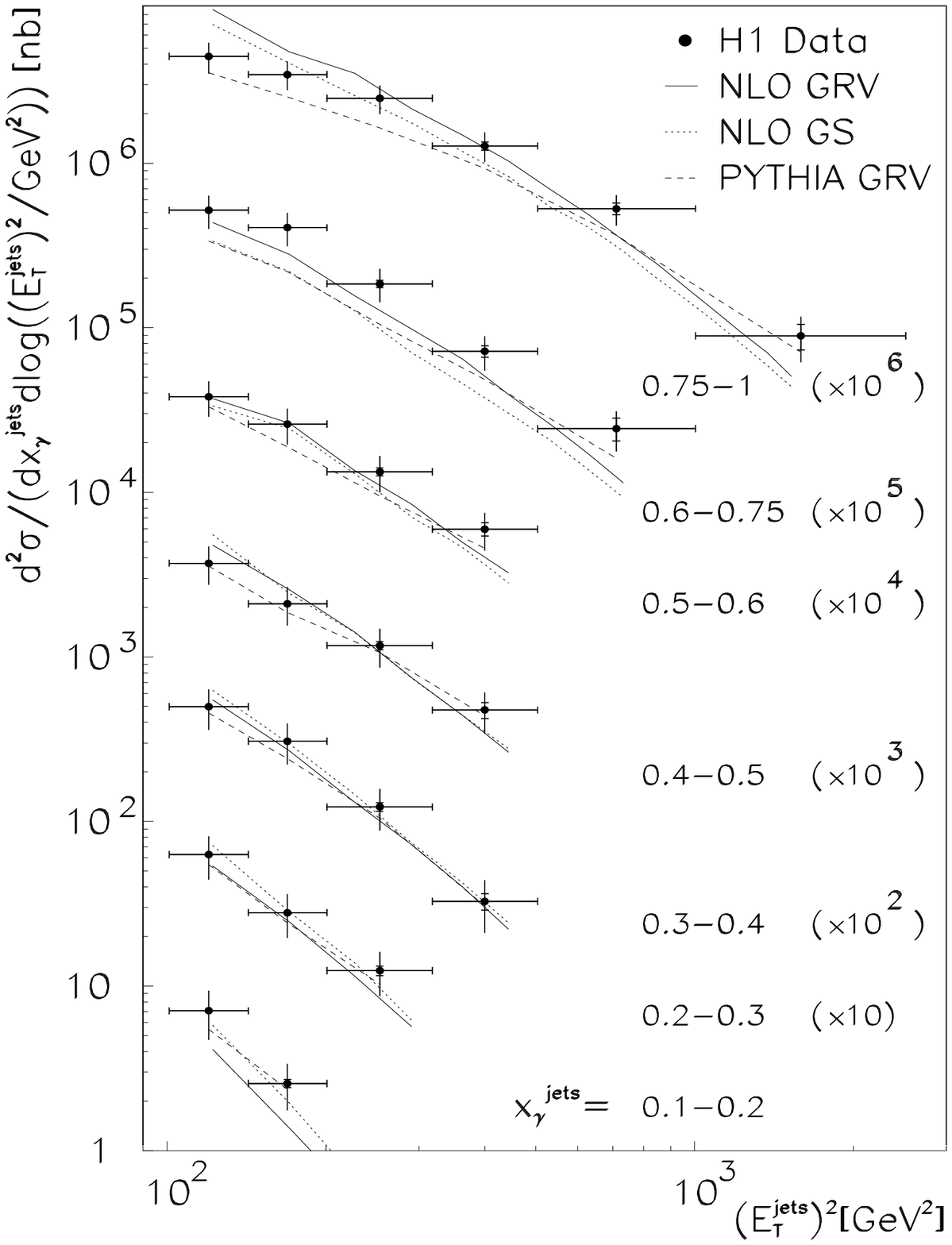,height=3.5in,width=2.5in,%
bbllx=10pt,bblly=69pt,bburx=460pt,bbury=660pt,clip=}
\put(-80,235) {a)}
\end{minipage}
\begin{minipage}{72mm}
\vspace*{0.2cm}
\begin{center} b)
\end{center}
\vspace*{0.3cm}
\hspace{-0.65cm}
\epsfig{file=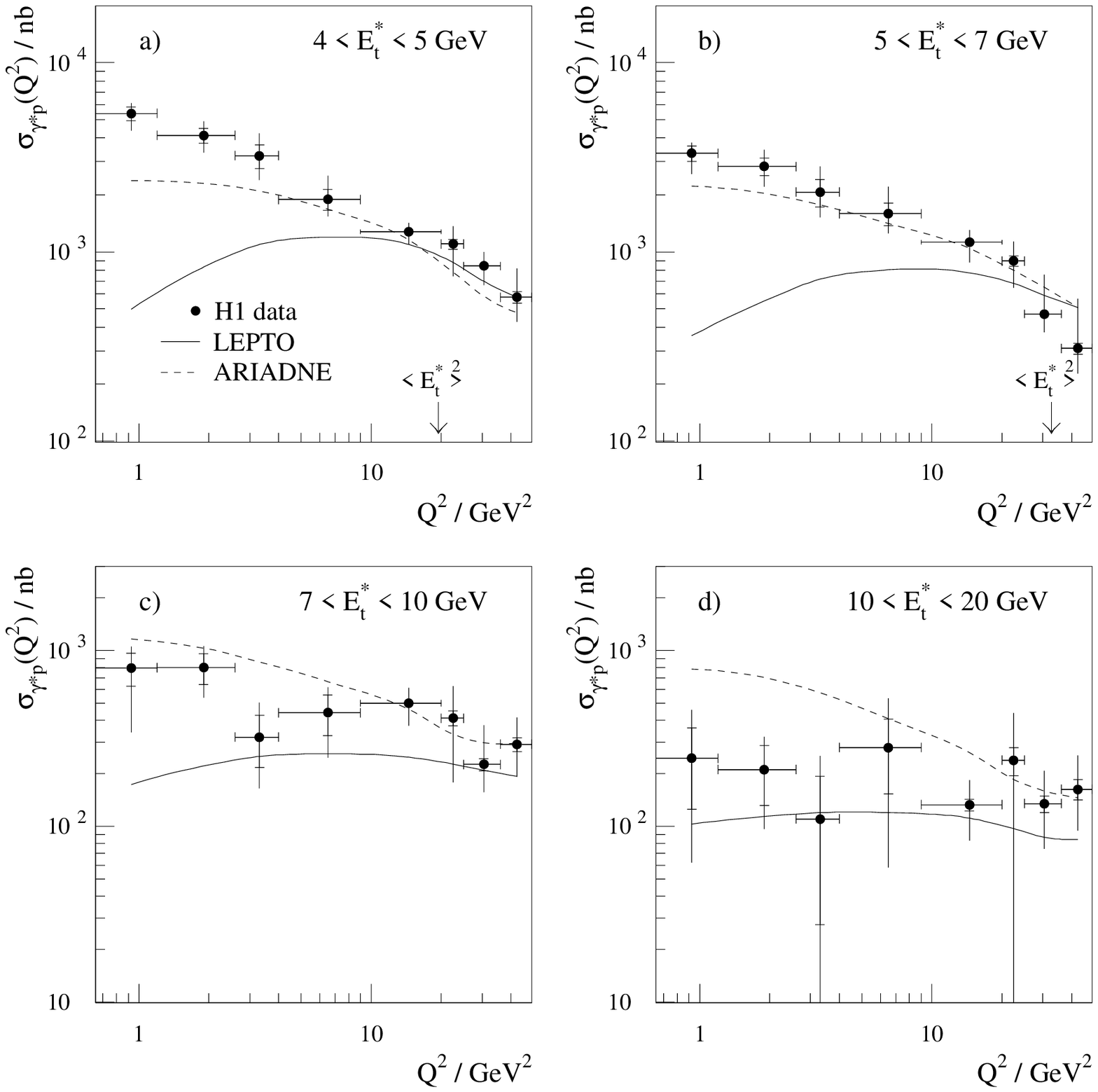,height=3in,width=3in,%
bbllx=13pt,bblly=16pt,bburx=512pt,bbury=514pt,clip=}
\end{minipage}

\vspace*{0.2cm} 
{\parbox[t]{14.5cm}
{\footnotesize Figure 1: a) The photoproduction di-jet cross section as a 
function of $(E_T^{jets})^2$ and $x_\gamma^{jets}$ in the range 
$0.2 < y < 0.83$. 
The data (full circles) are compared to the PYTHIA with the GRV-LO photon PDF
(dashed curve) and to analytical NLO calculations\cite{KK} with
GRV-HO (full line) and GS96 (dotted line) photon PDF.
b) The inclusive $\gamma^*p$ single-jet cross section as a
function of $Q^2$ in the range $0.3 < y < 0.6$. The data (full points) are
compared to LEPTO (solid line) and ARIADNE (dashed line).}
 
%\hfill
 
%\vspace*{-1.7cm}
%\begin{minipage}{75mm}
\hspace{2.5cm}
\epsfig{file=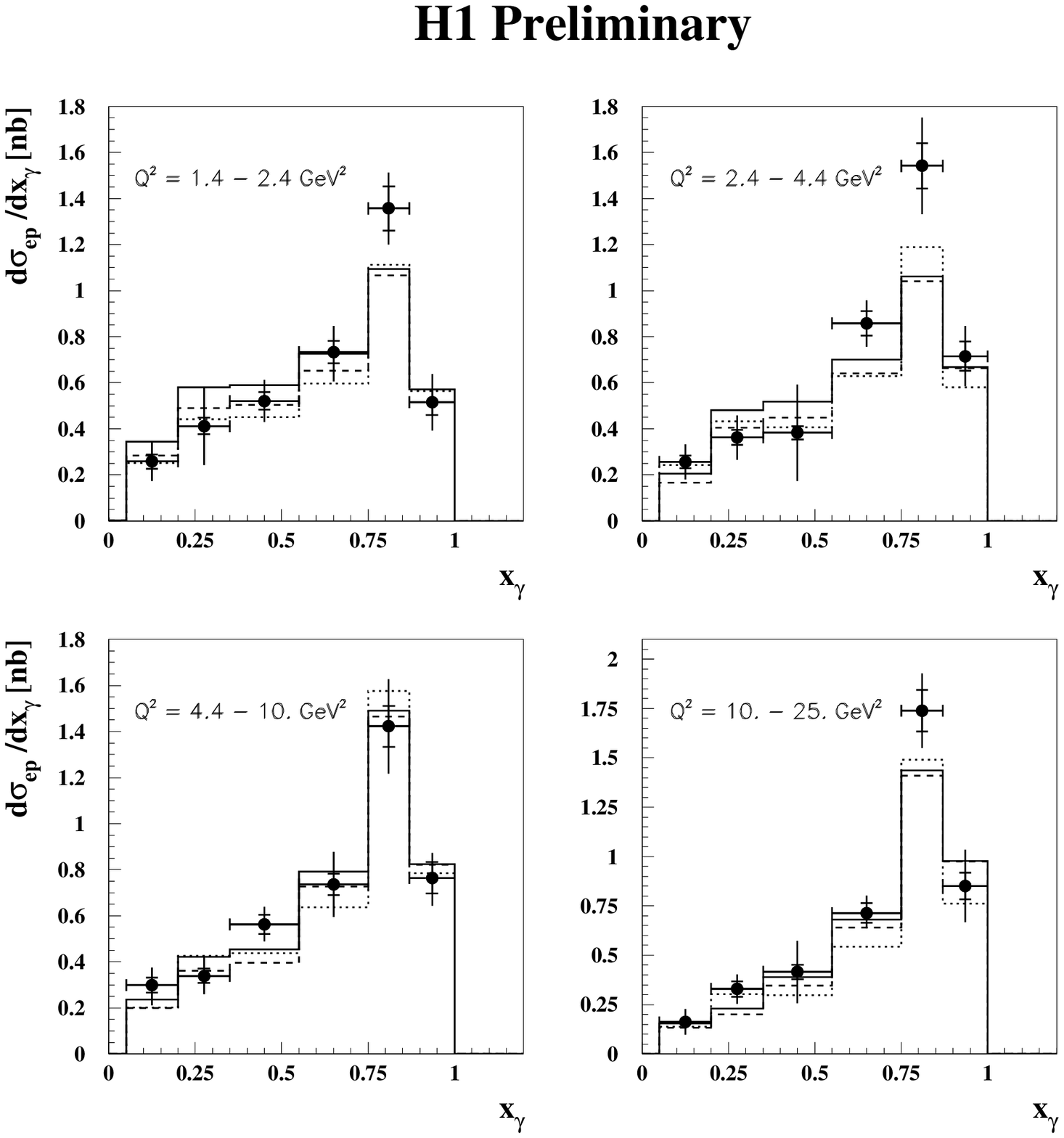,height=3.5in,width=3.5in,%
bbllx=29pt,bblly=154pt,bburx=526pt,bbury=710pt,clip=}
%\end{minipage}
%\begin{minipage}{45mm}
%\vspace*{2cm}
%H1 Preliminary \\ 
\vspace*{-0.1cm}

{\parbox[t]{14.5cm}
{\footnotesize Figure 2: The differential di-jet $ep$ cross sections as a 
function of $x_\gamma$ in four $Q^2$ intervals (points) compared to HERWIG 
with $\omega=0.1, 0.05$ GeV (full, dashed line resp.) and RAPGAP with 
$\omega=0.1$ GeV (dotted line) used in DG suppression function.}
%\vspace*{-2.1cm}
%\end{minipage}
 
\newpage
\vspace{-1cm}
\begin{minipage}{72mm}
\hspace*{0.7cm}
\epsfig{file=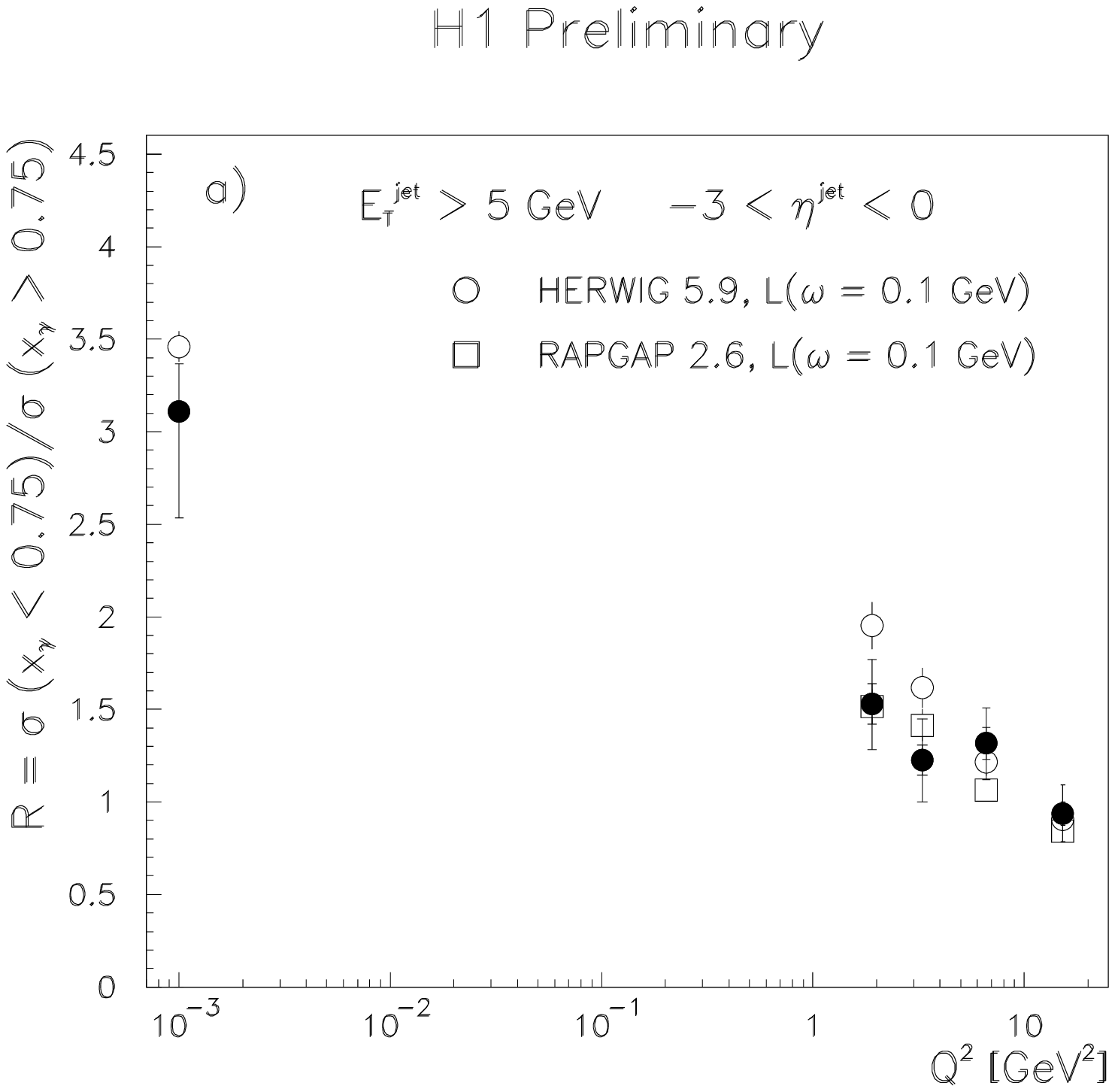,height=2in,width=2in,%
bbllx=56pt,bblly=117pt,bburx=482pt,bbury=554pt,clip=}
\end{minipage}
\begin{minipage}{72mm}
\vspace{0.8cm}
\epsfig{file=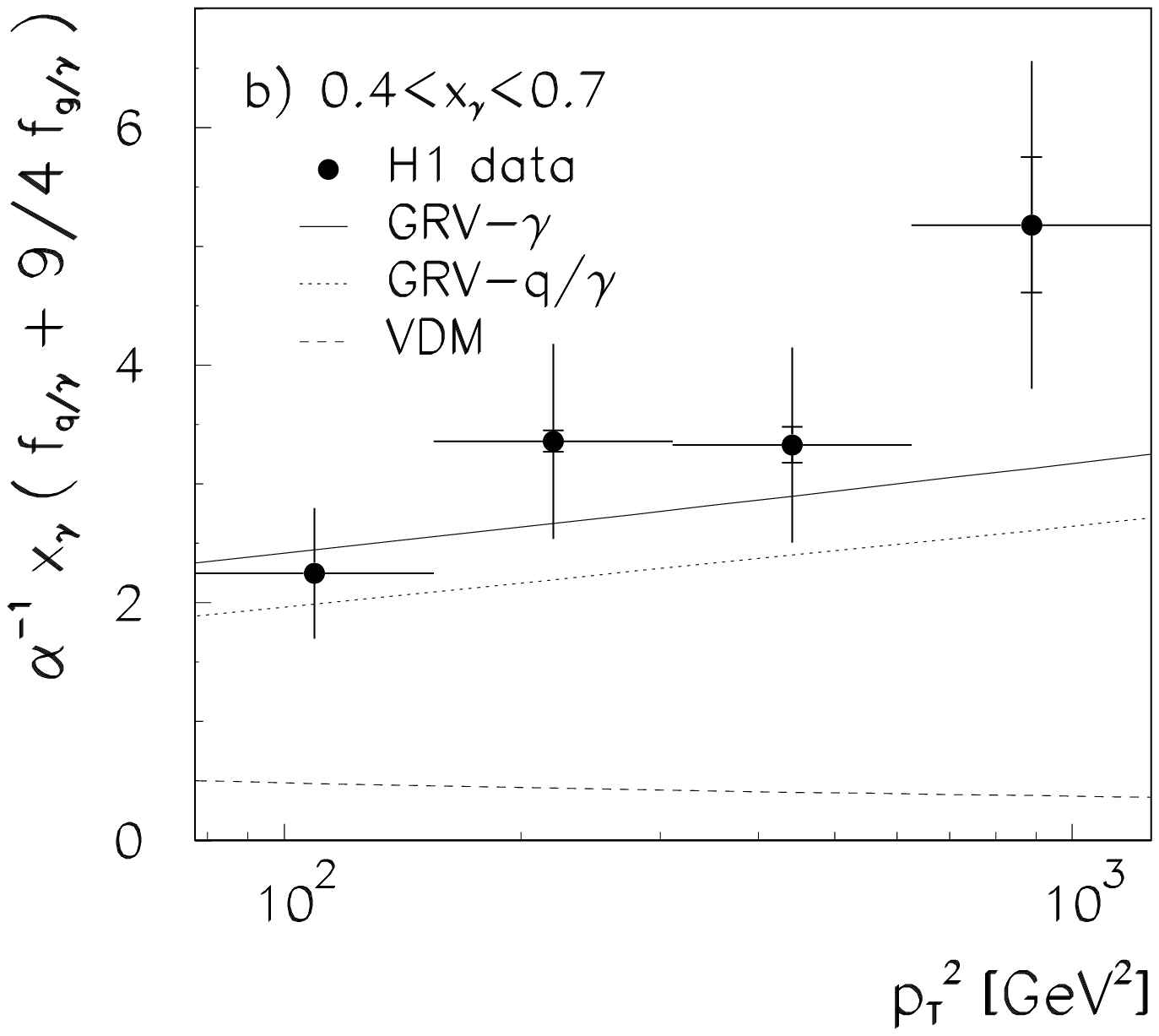,height=1.85in,%
bbllx=19pt,bblly=19pt,bburx=415pt,bbury=373pt,clip=}
\end{minipage}
\vspace{0.03cm}
 
{\parbox[t]{14.5cm}
{\footnotesize Figure 3: a) The ratio $\sigma_{res}/\sigma_{dir}$ as a 
function of $Q^2$. 
The data (full circles) are compared to the HERWIG (open circles) and RAPGAP 
model (open squares) with DG suppression for $\omega=0.1$ GeV.
b) The LO effective PDF of the photon $x_\gamma f_{\mathrm{eff}}^{\gamma}$ 
normalised to the $\alpha_{em}$ for $0.4 < x_\gamma < 0.7$
as a function of the squared parton momentum $p_T^2$. The data (points) are 
compared to the pointlike (full curve), VDM (dashed) and quark (dotted) 
component of the effective GRV-LO PDF of the photon.}
 
\begin{minipage}{145mm}
\vspace{0.4cm}
\hspace*{1.4cm}
\epsfig{file=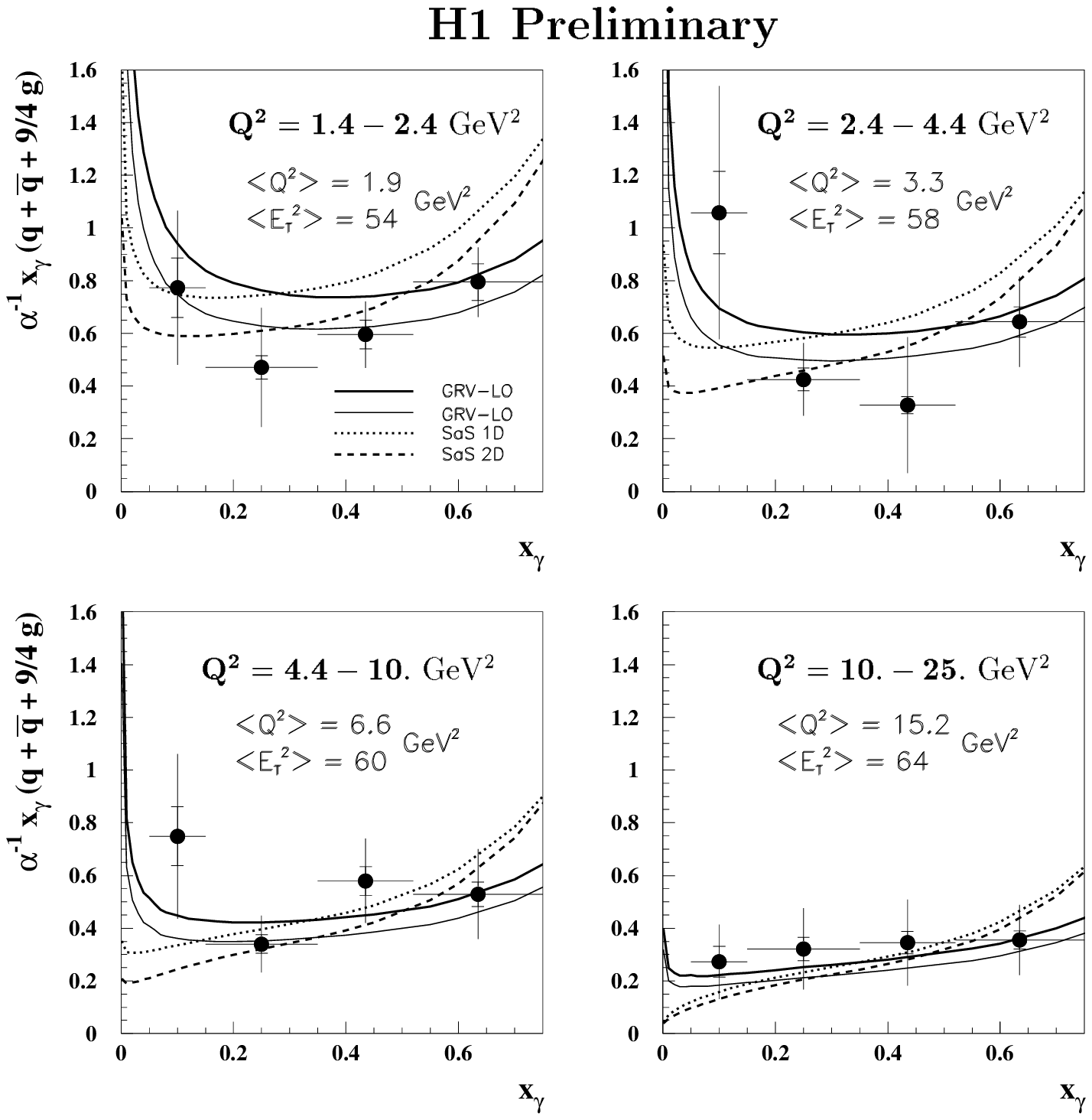,height=4in,width=4in,%
bbllx=73pt,bblly=221pt,bburx=480pt,bbury=640pt,clip=}
\end{minipage}
%\vspace{-2.cm}
\vspace{0.02cm}
 
{\parbox[t]{14.5cm}
{\footnotesize Figure 4: The LO effective parton distribution of the photon
$x_\gamma f_{\mathrm{eff}}^{\gamma}$, 
normalised to the $\alpha_{em}$, in four intervals of the 
photon virtuality $Q^2$. The data (points) are compared to the effective 
PDF of the photon using the GRV-LO parameterisation multiplied 
by DG suppression function with $\omega=0.1, 0.05$ 
GeV (full thick, full thin line resp.) and SaS1D (dotted line), SaS2D (dashed 
line) parameterisations.}

%c.  Use a hyphen (-) for compound words (e.g.
%`two-dimensional'), an en-dash (--) to link numbers, nouns or
%names (e.g. 220--240 Volts, electron--positron collisions,
%Einstein--Rosen--Podolsky paradox), and an em-dash (---) to link
%sentences or clauses---this is what we would regard as a
%`normal' dash.

\end{document}